\documentclass[a4paper,10pt,twoside]{article}
\usepackage{amsmath} %mathematical symbols etc.
\usepackage{amssymb} %mathematical symbols etc.
\usepackage{amsthm}  %for theorems, propositions, proofs etc.

\usepackage{fancyhdr}
\usepackage{enumitem}
\usepackage[utf8]{inputenc}
\usepackage{setspace}
\usepackage{graphicx}
\usepackage{textcomp}
\usepackage[retainorgcmds]{IEEEtrantools}
\usepackage{url}

\usepackage[a4paper,twoside, hmarginratio={1:1}]{geometry}

\newcommand{\oo}[1]{\overline{#1}}
\def\Rgamh{R_{\gamma\gamma\ h}}
\def\Rgam{R_{\gamma\gamma}}

\def\Rz{R_{ZZ}}

\def\RtauE{R_{\oo{\tau}\tau}^\text{exclusion}}
\def\tb{t_\beta}
\def\ma{m_{A^0}}
\def\beqn{\begin{eqnarray}}
\def\eeqn{\end{eqnarray}}

\begin{document}
\bibliographystyle{h-physrev}
\begin{titlepage}
\begin{center}

\vspace*{3cm}

{\Large {\bf BMSSM Higgses at 125 GeV }}
%\\

\vspace{8mm}

{F. Boudjema and  G. Drieu La Rochelle}\\

\vspace{4mm}

{ LAPTh$^\dagger$, Univ. de Savoie, CNRS, B.P.110,
Annecy-le-Vieux F-74941, France}

\vspace{10mm}

\today
\end{center}

\centerline{ {\bf Abstract} } \baselineskip=14pt \noindent
%%%%%%%%%%%%%%%%%%%%%%%%%%%%%%%%%%%%%%%%%%%%%%%%%%%%%%%%%%%%%%%%%%%%%

{\small
The BMSSM framework is an effective theory approach  that encapsulates a variety
of extensions beyond the MSSM with which it shares the same field content. The lightest
Higgs mass can be much heavier than in the MSSM without creating a tension with naturalness
or requiring superheavy stops. The phenomenology of the Higgs sector is at the same time much richer.
We critically review the properties of a Higgs with mass around 125 GeV in this model. In particular,
we investigate how the rates  in the important
inclusive $2\gamma$ channel, the $2\gamma + 2\; {\rm jets}$ and
the $ZZ \to 4l$ (and/or $WW$) can be enhanced or reduced compared to the standard model and what kind of correlations between these rates are possible. We consider both a
vanilla model where stops have moderate masses and do not mix and a model with large mixing and a light stop. We show that in both cases there are
scenarios that  lead to enhancements in these rates at a mass of 125 GeV corresponding either to the lightest Higgs or the heaviest CP-even Higgs of the model. In  all of these scenarios we study the prospects of finding other signatures either of the 125 GeV Higgs or those of the heavier Higgses. In most cases the $\bar\tau\tau$ channels are the most promising. Exclusion limits from the recent LHC Higgs searches are folded in our analyses while the tantalising  hints for a Higgs signal at 125 GeV are used as an example  of how to constrain the BMSSM and/or direct future searches.
}

\vspace*{\fill}

\vspace*{0.1cm} \rightline{LAPTh-013/12}

\vspace*{1cm}

$^\dagger${\small UMR 5108 du CNRS, associ\'ee  \`a l'Universit\'e
de Savoie.} \normalsize

\vspace*{2cm}

\end{titlepage}

\renewcommand{\topfraction}{0.85}
\renewcommand{\textfraction}{0.1}
\renewcommand{\floatpagefraction}{0.75}

\section{Introduction}
The latest upturn in the hunt for the Standard Model Higgs came at
the very end of 2011, when the ATLAS and CMS collaborations both
presented their analyses with the latest dataset (see
\cite{atlas_5fb,cms_5fb}) and showed that, besides an exclusion
limit that is driving the Higgs mass in a very thin region 115-131
GeV, there might be the possibility of a Higgs signal around
$m_h=125$ GeV. If these results are confirmed they will mark the
crowning of the Standard Model especially that this mass range is
in excellent agreement with the indirect limit from the global
electroweak precision measurements that test the inner working of
the model, shall we say the theory now, at the quantum and
renormalisable level. \\
\noindent With the fact that no new particle
outside the SM has been discovered, a lone elementary Higgs may
bring back the issue of naturalness. A known solution to this
problem is supersymmetry. No wonder that the first flurry of
articles after the announcement of a hint of a Higgs with mass
around $125$ GeV were from aficionados of supersymmetry. $125$ GeV
is an almost lucky strike for its minimal manifestation, the MSSM.
Almost lucky, because in the MSSM this value is on the heavy side,
requiring large values for the mass scale in the stop
sector\cite{Heinemeyer:2011aa,arbey_higgs_125,
carena_higgs_125,draper_higgs_125,cao_higgs_125,Feng:2011aa,Aparicio:2012iw,Desai:2012qy}.
This large scale brings back again the naturalness problem.
Extended models of
supersymmetry\cite{hall_higgs_125,Arvanitaki:2011ck} fare better
from this point of view, in particular one of the simplest
versions, namely the next-to-minimal version, the
NMSSM\cite{ellwanger_higgs_125,gunion_higgs_125,muhlleitner_higgs_125,cao_higgs_125},
can quite naturally provide a Higgs with $m_h=125$ GeV. These more
natural extended models allow also a richer phenomenology in the
Higgs sector than in the MSSM. For example, scrutinising the data
suggests that the signal corresponds to larger production rates
than in the SM in the $2\gamma$ channel. This is extremely
difficult to attain in the MSSM
\cite{carena_higgs_125,cao_higgs_125} where apart from the large
values for the scales in the stop sector to obtain the Higgs mass,
one needs the collaboration of staus with the mass of the lightest
stau as low as what is permitted by LEP but otherwise very large
parameters for other scales in this sector
also\cite{carena_higgs_125}. There is more flexibility with the
NMSSM\cite{ellwanger_higgs_125,cao_higgs_125,kang_higgs_125} when an increase in
the diphoton rate comes easily as the consequence of a drop in the
coupling of the Higgs to bottom quarks due to mixing with the
singlet component. Although we must stress that, considering the
significance of the results in the different channels and the two
experiments, one should take great care in drawing any hasty
conclusion, it must nonetheless be admitted that the hint of  a
signal is at the moment still compatible with a SM interpretation. \\

One could attempt to better fit the data within the most general
effective Lagrangian describing the Higgs interactions within an
extension of the
SM\cite{falkowski_higgs_125,contino_higgs_125,grojean_higgs_125}.
Fits to the anomalous Higgs couplings could then be performed.
This could be useful in constraining some underlying models but
this approach may not be as constraining  when the underlying
model tightly relates the different anomalous couplings of the
$125$ GeV Higgs that may also have a bearing on other (heavier or
lighter) Higgses that should be included in the picture. A case in
mind is the Higgs sector of the
BMSSM\cite{brignole,dine_bmssmhiggs_0707,antoniadis_bmssm_0708,antoniadis_bmssmhiggs_0806,ponton_susy_higgs_0809,
romagnoni_bmssmgoldstino_1111} which is a general effective
Lagrangian approach to the MSSM first introduced to address the
naturalness
problem\cite{espinosa_finetuning_2004,ross_finetuning_0903}.
Masses up to $250$ GeV can be naturally achieved for the lightest
Higgs of the
model\cite{carena_bmssmhiggs_0909,antoniadis_bmssmhiggs_0910,carena_bmssmhiggs_1005,
antoniadis_bmssmhiggs_1012,carena_bmssmhiggs_1111,gdlr_higgs_1112}.
Most recent analyses of the model taking into account ATLAS and
CMS data now only allow masses below $140$ GeV for the lightest
Higgs\cite{carena_bmssmhiggs_1111,gdlr_higgs_1112}. Naturally
there is no problem for the model to generate lightest Higgs
masses around $125$ GeV.  At the same time Higgses with this mass,
in this model, have a rich phenomenology with properties that can differ from those of the SM and the MSSM. It
is therefore very important to review the signatures of the
different manifestations of these Higgses. In particular it is
crucial to see which channels, in particular the most sensitive
ones: $\gamma \gamma$, $ZZ(4l)$, $\gamma \gamma \;+\;2\; {\rm
jets}$, in this mass range see their rates either enhanced or
reduced and study correlations between these rates.
This is the main purpose of this paper. \\

Apart from the profile of the Higgs with $m_h=125$ GeV in terms of
a signal in different channels, we take different constraints in
particular non observation of signals of other Higgses with the
present LHC luminosity.  We then take as an example a situation
where an enhancement in the inclusive diphoton channel and the
$ZZ\to 4l$ is confirmed and investigate what consequences on other
signals either of the same Higgs in other channels or other
heavier Higgses are to be expected when more data is collected. We
also review the role that the stop sector can play in the BMSSM.
In these models light stops can very easily give $m_h=125$ GeV,
however very light stops with large mixings can change in an
important way the correlations between the $\gamma \gamma$,
$ZZ(4l)$, $\gamma \gamma + 2 \;{\rm jets}$ channels. We also
investigate whether the $125$ GeV Higgs could correspond to  the
heaviest CP even Higgs. The article is organised as follows. In
section~\ref{bmssm_set_up} we briefly describe the BMSSM set up
and the main characteristics  of the Higgs that ensue, and we
define the parameter space and how the experimental data is
incorporated. In section~\ref{case_h}, we consider the case where
the lightest CP-even Higgs, $h$, has mass $m_h=125$ GeV. In
section~\ref{case_hh} we turn to a scenario where it is the
heaviest Higgs, $H$, that has mass $m_H=125$ GeV.
Section~\ref{sec_conclusions} collects our conclusions.

\section{BMSSM description}
\label{bmssm_set_up} The BMSSM is an effective theory that builds
upon the MSSM by the addition of higher order operators. This
means that it shares the same field content as the MSSM, in
particular in the Higgs sector the physical states are the 2
CP-even Higgses (the lightest $h$ and heaviest $H$), the CP-odd
$A^0$ and the charged Higgs $H^\pm$. The higher order operators
represent the effect of physics beyond the MSSM that has been
integrated out and is characterised by a mass scale $M$. The
effective theory is then an expansion in powers of $1/M$ of the
K\"ahler potential $K$ and the superpotential $W$. An exhaustive
set of leading order operators has been catalogued in
\cite{wudka_piriz_bmssm_1997} but what concerns us here are those
operators that have most impact, namely those that involve the
Higgs sector. Indeed, it is well known that the Higgs sector of
the MSSM is very much constrained and it is thanks to the
radiative corrections that the MSSM has survived so far. Likewise
perturbing a little through the introduction of these higher
operators changes the phenomenology of the MSSM quite drastically
and improves the naturalness argument. The set of leading order
operators, beyond the MSSM, is quite
restricted\cite{brignole,dine_bmssmhiggs_0707,antoniadis_bmssm_0708,
antoniadis_bmssmhiggs_0806,ponton_susy_higgs_0809}
\begin{IEEEeqnarray}{rCl}
W_{\text{eff}}&=&\zeta_1\frac{1}{M}\left(H_1.H_2\right)^2, \\
K_{\text{eff}}&=&a_1\frac{1}{M^2}\left(H_1^{\dag}e^{V_1}H_1\right)^2+a_2\frac{1}{M^2}\left(H_2^{\dag}e^{V_2}H_2\right)^2+a_3\frac{1}{M^2}\left(H_1^{\dag}e^{V_1}H_1\right)\left(H_2^{\dag}e^{V_2}H_2\right)\nonumber\\
&&+a_4\frac{1}{M^2}\bigl(H_1.H_2\bigr)\left(H_1^{\dag}.H_2^{\dag}\right)+\:\frac{1}{M^2}\left(a_5H_1^{\dag}e^{V_1}H_1+a_6H_2^{\dag}e^{V_2}H_2\right)\left(H_1.H_2+H_1^{\dag}.H_2^{\dag}\right).
\end{IEEEeqnarray}
where $H_1,H_2$ are the two Higgs superfields in the gauge basis,
assuming the hypercharges to be $Y_1=-1,Y_2=1$. The new parameters
$\zeta_1$, $a_i$ ($i=1..6$) are the remnant of the New Physics. In
order to account for supersymmetry breaking of these extra operators
the simplest way is to convert those effective coefficients into
spurions (see \cite{antoniadis_bmssmhiggs_0910}) :
\begin{align}
\zeta_1&\longrightarrow\zeta_{10}+\zeta_{11}m_s\theta^2,\\
a_i&\longrightarrow
a_{i0}+a_{i1}m_s\theta^2+a_{i1}^*m_s\overline{\theta}^2+a_{i2}m_s^2\overline{\theta}^2\theta^2.
\end{align} We have introduced the spurion breaking mass $m_s$ as
a book-keeping quantity so that all effective coefficients are
dimensionless. One finds that the leading corrections are of order
$m_s/M$ and $\mu/M$, where $\mu$ is the usual supersymmetric Higgs
mixing parameter. At order $1/M^2$ corrections are of order
$(m_s/M)^2$, $(\mu/M)^2$ and $(\mu m_s/M^2)$. There are also
corrections of order $v^2/M^2$, $v$ is the Standard Model vacuum
expectation value. We assume the underlying theory to be
approximately supersymmetric so that $m_s/M$ is small and the
effective expansion well behaved. In our analysis we take
\beqn
\frac{m_s}{M}=\frac{\mu}{M}=0.2, \quad \quad M=1.5{\rm TeV}.
\eeqn
Note that $\mu$ is (naturally) small, $\mu=m_s=300$ GeV. \\

We refer the reader to references
\cite{brignole,carena_bmssmhiggs_0909,antoniadis_bmssmhiggs_0910,gdlr_higgs_1112}
for the full derivations of the Higgs masses and couplings, in
particular how the cross sections and branching fractions, either
tree-level or loop induced, are computed\cite{gdlr_higgs_1112}.
Here we will just outline the main features. \\

With the coefficients ($\zeta, a$) in the range $-1,1$, masses for
the lightest Higgs up to $250$ GeV can be obtained. The largest
values are obtained for the smallest $\tb$ in the range $2-40$.
$\tb$ is the ratio of the vacuum expectation values of the Higgs
doublets. The coupling $VVh/H$ ($V=W^\pm,Z$) can not be larger
than 1 but can be very much reduced. The implementation imposes a
custodial symmetry, then the $WWh$ and $ZZh$ are to a very good
approximation equal. $t\bar t h$ is not much affected especially
compared to $b \bar b h$ that can be either greatly reduced or
enhanced. This has important consequences for gluon fusion and the
branching fraction of the Higgs into photons and $ZZ$ in the range
of interest around $125$ GeV.

\begin{figure}[h!]
\begin{center}
\includegraphics[scale=0.3,trim=0 0 0 0,clip=true]{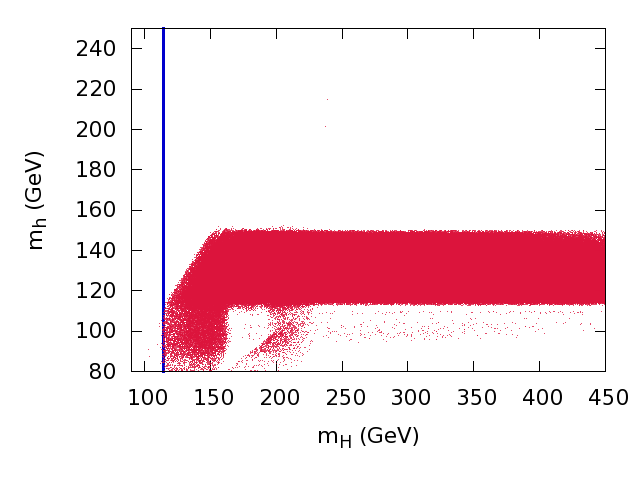}
\end{center}
\caption{\label{fig:pre_signal} {\em We plot here the allowed
region in the $m_H,m_h$ plane with the $2$ fb$^{-1}$ dataset from
LHC. For more details, see \cite{gdlr_higgs_1112}.}}
\end{figure}

LEP, Tevatron and in particular the LHC data have put dramatic
constraints on the model\cite{gdlr_higgs_1112}. Taking into
account the status of the searches at ATLAS and CMS with a dataset
up to $2.3$ fb$^{-1}$ \cite{atlas_cms_lp_comb} (the parameter space and the experimental
constraint being discussed in the following part), the lightest
higgs, $h$ can not be heavier than $140$ GeV, see
fig.~\ref{fig:pre_signal}. Note that the data allow for the
heaviest CP even Higgs, $H$, to be quite light with $m_H=125$ GeV,
with a light Higgs, $h$, escaping the LEP bounds. There is
therefore the possibility that the the hint for an excess at
$125$ GeV may be due to $H$ with $h$ being more elusive, a
possibility that we will address.
\\

The production channels are also modified. Gluon fusion
($gg\rightarrow h$) can be either greatly enhanced or suppressed :
when normalized to the  SM expectation its value ranges from 0.1
to higher than 5. Vector boson fusion, $VV \to h/H$,  on the other
hand can only be reduced. The $b$ fusion ($b\oo{b}\rightarrow h$)
can  be greatly enhanced or reduced with respect to the SM
prediction. It is important to note that the production mechanisms
that contribute to the same inclusive channel are not rescaled in
the same way : hence the differential distributions of the BMSSM
Higgs production will not be exactly the same  as the SM ones
since each production mode can lead to different kinematic
properties. This remark should be kept in mind when exploiting SM
limits based on inclusive cross section. We discussed this issue
at some length and gave recommendations in\cite{gdlr_higgs_1112}.

\begin{figure}[h!]
\begin{center}
\includegraphics[scale=0.5,trim=0 0 0 0,clip=true]{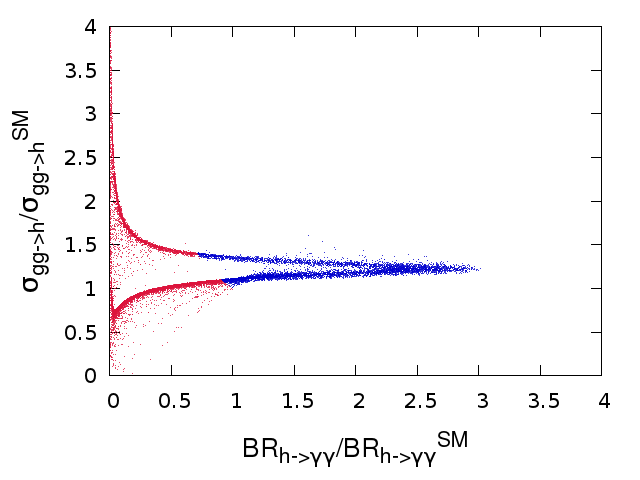}
\end{center}
\caption{\label{fig:glu_b} {\em We show here the gluon fusion
normalised to the SM expectation as a function of the normalised
branching ratio to photons in the BMSSM, with $122<m_h<128$ (GeV),
before applying collider constraints. The effect of the stops is
negligible here.}}
\end{figure}

For the decays of the Higges, the branching ratios can also be
enhanced or suppressed, however these changes are not independent
from how the production rates are affected. Since we are
interested in the mass range around $m_h=125$ GeV, we plot in figure
\ref{fig:glu_b} the normalised value of the gluon fusion as a
function of the normalised branching ratio into photons for $h$,
with red (blue) points for $\Rgam<1$ ($\Rgam>1$) in the range
$122<m_h<128$ (GeV). In this scenario the stop sector has little
impact. The ratio $\Rgam$, to be better specified later, corresponds to a
good approximation to the product of the normalised gluon fusion
and normalised branching ratio into photons. As can be seen,
enhancing the branching ratio into photons has a strong
consequence on gluon fusion. An enhanced branching fraction into
photons forces $\sigma_{g g \to h}$ to have an almost SM rate. For the
highest value of $Br(h\to \gamma \gamma)/Br^{SM}(h\to \gamma
\gamma)$, $\sigma_{g g \to h}/\sigma_{g g \to h}^{SM}\sim 1.2$.
This slight increase , together with the structure of the correlation (the two
branches) see fig.~\ref{fig:glu_b},  can be understood as follows. In this set-up $gg \to h$
is driven mostly by the top contribution that is practically
standard model like and the bottom contribution that can vary
greatly. In the SM, the small $b$ contribution interferes
destructively with the dominant top contribution. If
now we parameterise the strength of the $hb\bar b$ in the BMSSM as
$h^{\rm BMSSM}b \bar b= x_b h^{\rm SM}b \bar b$, then we
approximately have
\beqn
\frac{\sigma_{g g \to h}}{\sigma_{g g \to
h}^{SM}}=\frac{(1-\epsilon_{bt} x_b)^2}{(1-\epsilon_{bt})^2} \quad
\quad 0<\epsilon_{bt} \ll 1,
\eeqn
$\epsilon_{bt}$ is the relative contribution of the bottom
compared to the top, at the amplitude level. Largest values of $Br(h\to \gamma
\gamma)/Br^{SM}(h\to \gamma \gamma)$ are due to a much reduced  $h \to b \bar
b$ width and hence small $|x_b|$. Then for $|x_b|<1$

\beqn
\frac{\sigma_{g g \to h}}{\sigma_{g g \to
h}^{SM}}=\frac{(1-\epsilon_{bt} x_b)^2}{(1-\epsilon_{bt})^2}
&\sim& 1+ 2 \epsilon_{bt} (1-x_b) > 1,\nonumber \\
& \simeq & 1+ 2 \epsilon_{bt} \quad x_b \to 0.
\eeqn
which is the contribution of the bottom contribution. The bell
shape is a reflection of the quadratic dependence
$(1-\epsilon_{bt} x_b)^2$. We should keep this characteristic in
mind for the rest of the analysis. Therefore despite what might
seem to be a relative freedom with the new parameters introduced
by the BMSSM, some predictions are rather constrained. Let us add a comment about the reduction
of the $h b \bar b$ coupling. With $m_h \sim 125$ GeV, even a relatively large reduction which is accompanied by an
almost similar reduction of the $h \oo{\tau}\tau$ coupling, might still not affect so drastically the $Br (h \to b \bar b)$ nor the $Br (h \to \oo{\tau}\tau)$
since $\Gamma (h \to b \bar b)$ might still remain dominant.

\subsection{The parameter space}
$\tb$ and $\ma$ (the mass of the pseudoscalar Higgs) are varied in
the following range.
$$\tb\in[2,40], \qquad\ma\in[50,450].$$

Considering the impact of the third family on Higgs physics, we
allow some flexibility in the stop sector. Two implementations
will be considered.
\begin{itemize}
\item Model A: A no mixing scenario, where all soft masses of the third generation
squarks are set to $M_{{u3}_R}=M_{{d3}_R}=M_{Q_3}=400$ GeV with no
mixing, in particular the stop mixing parameter $A_t=0$. For these
values the masses $m_{\tilde{t}_1},m_{\tilde{t}_2}$ are around
$400$ GeV and the mixing angle vanishes. This is taken as a
standard case, where stops are not too heavy and in the set up of
the BMSSM their effect is not so important.
\item Model B: A maximal
mixing scenario where one of the stop is light
$m_{\tilde{t}_1}=200$ GeV. We will take
$m_{\tilde{\tilde{t}}_2}\in[300,1000]$ (GeV) and $\sin
2\theta_{\tilde t}=-1$. The heaviest stop mass is taken as a free
parameter. This will have important consequences in the production
of the Higgses and their decays. Note that, in a generic model, a
$200$ GeV stop can still escape all current collider limits.
\end{itemize}

For the rest of the MSSM parameter space, the following values
have been used : all soft masses are set to $M_{{\rm soft}}=1$ TeV
(except for the third generation), $\mu$ and $M_2$ (the $SU(2)$
gaugino mass) are set to 300 GeV, $M_1$ (the $U(1)$ bino mass)is
fixed by the universal gaugino mass relation
$M_1=\frac{5}{3}\tan^2 \theta_W M_2 \simeq M_2/2$, and $M_3=800$
(the $SU(3)$ gaugino mass) GeV, $\cos^2\theta_W=M_W^2/M_Z^2$. All
trilinear couplings are set to 0 (except $A_t$).\\

For the purpose of this paper this last set of parameters has
practically no impact, so we could have taken lower values of
$M_\text{soft}$ and $M_3$ compared to the new scale $M=1.5$ TeV. In
scenario B, the largest value of the heaviest stop
$m_{\tilde{t}_2}=1$ TeV that we allow in the scan should be
regarded an extreme example, not only from the point of view of
naturalness but also because it is not far heavier than the new
scale $M=1.5$ TeV. This said we must emphasise that most of our
study is done with $m_{\tilde{t}_2}=600$ GeV. Furthermore the
heavy scale $M$ can be enhanced with little change to our results
provided one keeps fixed the  ratios $m_s/M,\mu/M$.\\

The effective coefficients, ($\zeta,a$) will be varied in the
range $[-1,1]$. The following constraints were applied on each
point, in order to constrain the effective parameter space (see
\cite{gdlr_higgs_1112} for more details) :
\begin{itemize}
\item Perturbativity check. We check  that
$1/M^3$ contributions to $m_h$ are small enough. Indeed one must
make sure that the perturbative expansion in $1/M$ is under
control. This means that on top of all experimental constraints
that we take into account, we also check that our points exhibit a
correct effective expansion by keeping only points where the third
order ({\it i.e.} $1/M^3$ terms) contribution to the light Higgs
mass is within $10\%$ of its value.
\item
Electroweak Precision tests. Since the effective coefficients have
a non vanishing contribution to oblique parameters, we verify the
consistency with the electroweak precision measurements.
\item LEP
and Tevatron Higgs searches. This applies to  all Higgses :
$h,H,A^0,H^+$, including top decays to $H^\pm$.
\item ATLAS and CMS Higgs searches. The list of channels that we
take into account and how these are exploited within the BMSSM is
detailed in\cite{gdlr_higgs_1112}.  As concerns the exclusion
limits exploiting the $\oo{\tau}\tau$ channels we had also included
in\cite{gdlr_higgs_1112} the MSSM analysis of ATLAS and CMS.
Concerning the mass range $122-128$ GeV and an  interpretation in
terms of a signal, we detail our approach  in section
\ref{input_lhc_hint}.
\end{itemize}

In this work we do not include constraints based on flavour
physics ($(g-2)_\mu$, $B_s \to \mu^+ \mu^-$,..) and dark matter
(relic density) constraints. For the former, this will introduce
some extra model dependence from extra operators in the BMSSM,
outside the Higgs sector. For the relic density it is known that
the prediction can change drastically if we change the
cosmological model. These issues may obscure the conclusions on
the interpretation of the possible Higgs signal. We will certainly
come back to these effects in future investigations.

\subsection{Input from the LHC}
\label{input_lhc_hint}
In order to use the results from the ATLAS and CMS collaborations,
we have used the following ratios
\begin{equation}
\label{RXX} R_{XX}=\frac{\sigma_{pp\rightarrow h\rightarrow
XX}}{\sigma_{pp\rightarrow h\rightarrow
XX}^{SM}}\qquad\text{and}\qquad
R_{XX}^\text{exclusion}=\frac{\sigma_{pp\rightarrow H\rightarrow
XX}}{\sigma_{pp\rightarrow H \rightarrow XX}^{\text{excluded
95\%}}},
\end{equation} where $XX$ denotes a particular final state
(say the inclusive $2\gamma$). $\sigma^\text{excluded 95\%}$
stands for the 95\% C.L. excluded cross-section reported by the
collaborations. In practice the $R_{XX}$ will be used in the
signal case, to compare with the best fit $\hat{\mu}$ -- of the so
called signal strength $\mu$-- given by the experiments. In
eq.~\ref{RXX}, $h$ in the BMSSM  will refer either to the lightest
or heaviest CP-even Higgs. $R_{XX}^\text{exclusion}$ will be used
in the no-signal case as a measure of the sensitivity of the
search, here $H$ stands for all Higgses not contributing to a signal
in the mass range $122-128$ GeV. It will also be shown as a measure
of the luminosity needed to see the effect of a particular Higgs
in a certain channel in the future. For $R_{XX}$ the most
important channels are the inclusive $2\gamma$, $ZZ \to 4l$ (ATLAS
and CMS) and the exclusive $2\gamma +2 jets$ (CMS). For
$R_{XX}^\text{exclusion}$ we use the full dataset across all
Higgs masses covered by the experiments with the current collected
luminosity of $4.9{\rm fb}^{-1}$. As we said
$R_{XX}^\text{exclusion}$ gives a measure of the luminosity needed
to uncover a signal in a new channel. The current
$R_{XX}^\text{exclusion}$ is based on a collected luminosity of
about $4.9 {\rm fb}^{-1}$ for the LHC running at $7$ TeV. To
extract from the plots for $R_{XX}^\text{exclusion}$ that we will
show the approximate luminosity that will be needed to uncover a
potential signal, one can base the rough estimate on a simple
rescaling of the luminosity. For example in the plots we will
show, a channel with $R_{XX}^\text{exclusion}=0.1$ will require a
luminosity of about $500 {\rm fb}^{-1}$ to be observed. With a
luminosity of $30 {\rm fb}^{-1}$ only those channels with
$R_{XX}^\text{exclusion}>0.4$ may be accessible.

We decided not to carry out a fit of the model to the data (as was
done in \cite{grojean_higgs_125,contino_higgs_125} among others)
for the reason that so far the signals from both collaborations
are not so easy to reconcile and also because in view of the
quality of the results this exercise is far too premature. We
chose instead to focus on assessing the limits on the flexibility
of the BMSSM with a Higgs in the range $122< m_h < 128$ GeV,  a
range broad enough that it takes into account the hints from both
experiments including the uncertainty on the mass in each
experiment. We leave a complete computation of the compatibility
of the model with the data for the future, hopefully more precise,
set of data. By using those ratios instead of computing the
complete likelihood function (as was tried in
\cite{contino_higgs_125}), we have implicitly assumed some
approximations that are discussed  in \cite{gdlr_higgs_1112}. In
the inclusive channels, we neglect for instance the effect of
acceptance and efficiency cuts due to the change in the relative
contribution of the different production modes (gluon fusion,
vector boson fusion and associated production, $b$ quark fusion)
compared to the Standard Model. In other words, we are taking
ratios of inclusive cross-sections. For analyses that are truly
inclusive, such as the $h\rightarrow ZZ\rightarrow 4l$ there is
hardly any difference (see \cite{gdlr_higgs_1112}), but the
situation changes when channels are divided in subchannels with
different final states. The $h/H\rightarrow WW +0/1/2\text{ jets}$
comes to mind. The difficulty is that each subchannel that
contributes to the inclusive cross section has a different
efficiency and the fact that  most often the rates for the
different channels in the BMSSM are not rescaled by the same
factor. The reason for these approximations is that most of the
experimental quantities, such as the efficiency and the acceptance
for each production mode or the cross-section limits for each
subchannels, are so far unavailable, forbidding hence an exact
interpretation. So we reiterate our recommendation that
efficiencies be provided (see the Les Houches Recommendations for the Presentation of LHC results \cite{les_houches_reco_1203} for a detailed discussion on the subject.). However, for the time being those
approximations do not prevent a qualitative survey of the would be
Higgs signal properties.\\
Having said that, CMS\cite{cms_5fb_gamgam} has most recently
provided data for a more exclusive observable, $2\gamma +2{\rm
jets}$. The latter is more sensitive to the production through
vector boson fusion, with the Higgs subsequently decaying into 2
photons. We take this observable into account. The contribution of
the $gg \to h$ to this channel is however not negligible.
Fortunately the CMS collaboration\cite{cms_5fb_gamgam} does
provide the overall acceptance/efficiency (product of the two) for
each of the important channels (this applies also for the BMSSM),
albeit for a single Higgs mass, $m_h=120$ GeV. The overall
acceptance is quoted as $0.15$ for production through Vector Boson
Fusion (VBF)  and $0.005$ for gluon fusion. We therefore simulate
the ratio $R_{\gamma\gamma+\text{2 jets}}$ as
\begin{equation} R_{\gamma\gamma+\text{2
jets}}=\frac{0.15\,\sigma_{\text{VBF}}+0.005\,\sigma_{gg\rightarrow
h}}{0.15\,\sigma_{\text{VBF}}^{SM}+0.005\,\sigma_{gg\rightarrow
h}^{SM}}\ \times\ \frac{BR_{\gamma\gamma}}{BR_{\gamma\gamma}^{SM}}
\end{equation}
We checked that this parametrisation of
$\sigma_{\gamma\gamma+\text{2 jets}}$ when folded in with the SM
cross sections for the LHC at $7$ TeV\cite{LHC_Higgs_cs2} and
taking into account the luminosity quoted by CMS reproduced quite
exactly the number of selected events given by CMS
\cite{cms_5fb_gamgam}. We assume that this parametrisation that
was verified to be excellent for $m_h=120$ GeV still holds to a
very good degree in the range $122 < m_h < 128$ GeV.

In order to determine if a Higgs is excluded, we use a quadrature
combination of all $R^\text{exclusion}$ from all channels of both
experiments and checked if the result was below 1. We apply the
test on all Higgses ($h,H,A^0,H^+$), adding ratios when the Higgs
bosons are degenerate, before determining whether the parameter
point is allowed.

The signal condition we will require is the following : first we
require a Higgs (heavy or light) boson in the range $122-128$ GeV.
Then, we discard any exclusion limit on this Higgs in the
$\gamma\gamma$, $ZZ$ and $WW$ channels (since we want it to be a
signal in those channels). We apply on all other channels and on
all other Higgses the latest exclusion limits, obtained with the
$4.9$ fb$^{-1}$ dataset.

\subsection{Signal features, data}
The data that is most indicative of a possible signal is the
following (uncertainties correspond to the $1\sigma$ band)
\renewcommand{\labelitemi}{$\star$}
\renewcommand{\labelitemii}{$\bullet$}
\begin{itemize}
\item ATLAS\cite{atlas_5fb_gamgam}:\\
The ATLAS collaboration records a combined (all channels) signal
strength of $1.5$~${}_{-0.5}^{+0.6}$ at $m_h=126$ GeV. It may be
considered as most revealing in channels with best resolutions on
the Higgs mass:
\begin{itemize}
\item The inclusive $\gamma\gamma$ channel  where the signal
strength is $2^{+0.9}_{-0.8}$
\item $ZZ \to 4l$ channel  where the
signal strengths is $1.2^{+1.2}_{-0.8}$ compatible with $WW \to ll
\nu \nu$, though the $WW$ channel  has a worse mass resolution.
\end{itemize}
\item CMS collaboration\cite{cms_5fb} reports a combined signal strength of $1.2^{+0.3}_{-0.4}$ at $m_h=124$ GeV
\begin{itemize}
\item In the $\gamma \gamma$
channels, the first CMS release with 4.9 fb$^{-1}$ was based on an
analysis with four subchannels that gave a signal strength of $1.7\pm0.8$ at $m_h=123.5$
GeV (see ref (\cite{cms_5fb_gamgam_b})). The updated release added
a dijet-tagged subchannel $\gamma\gamma+\text{ 2 jets}$ yielding by itself a signal
strength of $3.8^{+2.4}_{-1.8}$. The combination of the five subchannels yield a signal strength of $2.1^{+0.8}_{-0.7}$ (\cite{cms_5fb_gamgam}).

\item For the $ZZ
\to 4l$, $0.5^{+1.0}_{-0.7}$. Note that the mean is low, moreover
the mean value for $m_h$ is at 126 GeV.
\item the $b\bar b$ and
$\tau^+ \tau^-$ channels analysed by CMS\cite{cms_5fb_gamgam_b} in the mass range $122-128$ GeV have so much uncertainty
that they are of little use in the present analysis.
\end{itemize}
\end{itemize}

Let us take breath and emphasise again that there is still much
uncertainty in these results, some of which may not help in
drawing a  coherent picture, expect perhaps in the $\gamma \gamma$
channel. The signal strengths are compatible with a Standard Model
Higgs, however it is tempting and in any case educative to
entertain the idea that some non standard Higgs scenario is
emerging. What is very interesting is that the different channels
and subchannels will allow, when measured with better precision,
to discriminate between different models and implementations of
the BMSSM. Most probably a first step in this discrimination in
this mass range will be performed with $\gamma \gamma, VV, \gamma
\gamma + 2\text{jets}$ perhaps also with the incorporation of the
$\oo{\tau}\tau$ channel. In the case of a multi-Higgs system this will
be done in parallel with searches for other Higgses. In the rest
of the paper we will investigate what kind of correlations between
these observables are possible within the BMSSM, for example
whether enhancements in all channels are possible.

\section{$h$ as a signal in the $122-128$ GeV window}
\label{case_h}
\subsection{Model A: No light stops, no mixing in the stop sector}
%\label{bmssm_set_up}
\begin{figure}[h!]
\begin{center}
\includegraphics[scale=0.3,trim=0 0 0 0,clip=true]{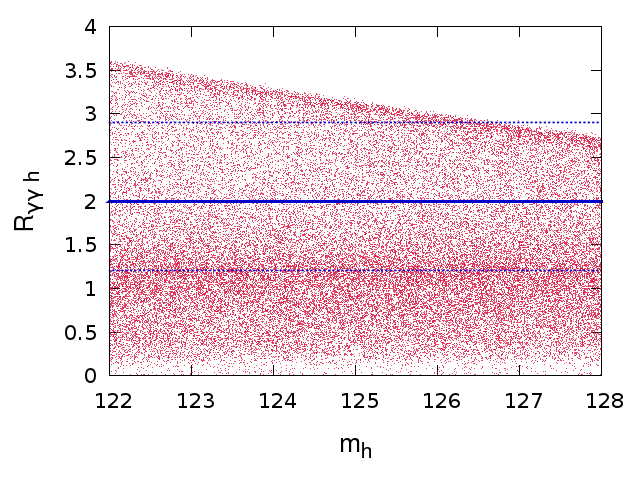}
\end{center}
\caption{\label{fig:h_Rgam} {\em Allowed region in the plane
$m_h,\Rgam$. The blue line represents the ATLAS best fit for the
signal strength, and the dotted lines are the one sigma deviations
from this value in model A.}}
\end{figure}
\begin{figure}[h!]
\begin{center}
\begin{tabular}{cc}
\includegraphics[scale=0.3,trim=0 0 0 0,clip=true]{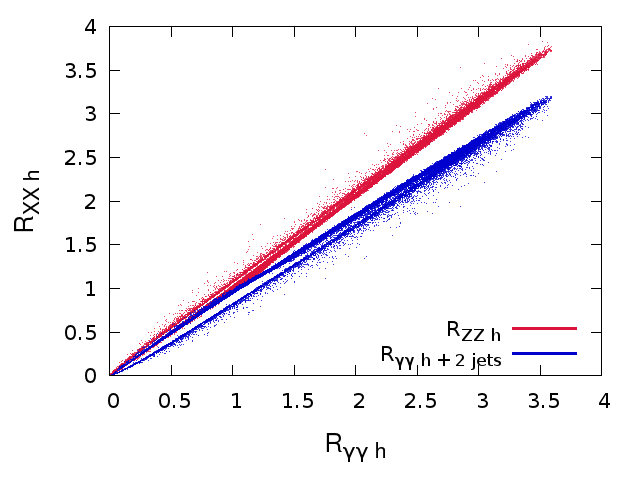}&
\includegraphics[scale=0.3,trim=0 0 0 0,clip=true]{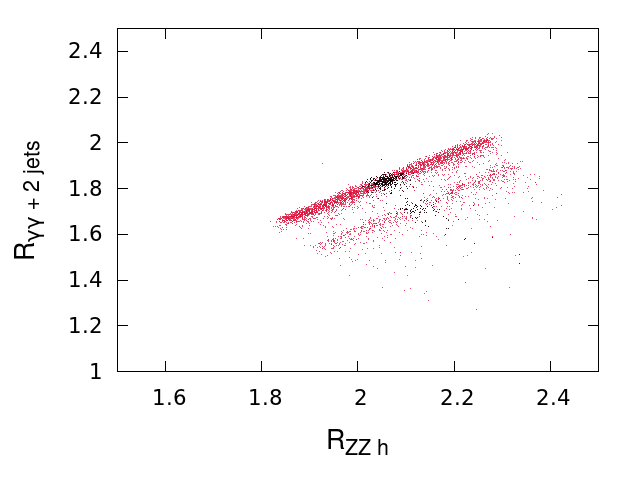}
\end{tabular}
\end{center}
\caption{\label{fig:nostop} {\em Left panel: correlation between
$\Rgam$, $\Rz$ and $R_{\gamma\gamma+\text{2 jets}}$ for $122< m_h
< 128$ GeV. Right panel: Imposing $\Rgamh=2.0\pm10\%$ (points in
red) and  $\Rgamh=2.0\pm1\%$ (points in black) we show the
correlation in the plane $\Rz$ and $R_{\gamma\gamma+\text{2
jets}}$. Both figures are for model A.}}
\end{figure}

Fig.~\ref{fig:h_Rgam} shows that with the current data, the BMSSM
yields a production rate in the inclusive $pp\rightarrow
h\rightarrow\gamma\gamma$ that can be quite small (as small as
$0.1$), and hence unobservable with the current luminosity or in
the very near future. More interestingly there is however no
difficulty finding a signal in this channel that is up to $3.5$
times that of the SM. There is a very strong correlation with the
signatures in the other promising channels: $VV \equiv ZZ \to 4l$
and the $2\gamma \; + \;2 \; jets$, see fig \ref{fig:nostop}. With
small differences we have $R_{\gamma\gamma} \simeq R_{ZZ} \sim
R_{\gamma\gamma+\text{2 jets}}$. Rates above those of the SM are
mostly driven by reduction in the width of to $b \bar b$ which
increases all channels. This is trivially seen for the $2\gamma$
{\it versus} $ZZ$ channel. In the case of the
$\gamma\gamma$/$\gamma\gamma+\text{2 jets}$ correlation, when the
rates are above those of the SM, the inclusive channel is higher
by 20\% or so, this (and the appearance of two branches) rests on
the same argument that we put forth in section~\ref{bmssm_set_up}
about the contribution of the $b$ quarks. Therefore  a
configuration with $R_{ZZ \to 4l}=1, R_{\gamma \gamma}=2,
R_{\gamma \gamma \; + \; 2 \; jets}=3$ is very much disfavoured in
Model A.

It is important to stress that the characteristics we find in
these scenarios occur for all values $\tb$, even if statistically,
with a simple scan, the population with smaller $\tb$ is larger.

\subsection{Model B: a light stop with large mixing in the stop sector}
It has been known for some
time\cite{Djouadi:higgsstopreduce,boudjema_light_stop} that, within
the MSSM, light stops endowed with a large mixing can drastically
reduce the $gg$ induced production. Even if this is accompanied by
an increase in the decay width to photons, the combined effect can
be a large drop in $gg \to h \to \gamma \gamma$. This effect is
encapsulated in the coupling of the stops to the Higgs. The
coupling of the lightest stop, $\tilde{t}_1$,
$g_{h\tilde{t}_1\tilde{t}_1}$  writes in the large $\ma$ limit
\begin{equation}
\label{eq:ght1t1}
 g_{h\tilde{t}_1\tilde{t}_1}\simeq\frac{g}{M_W}\left(\sin^2(2\theta_{\tilde{t}})\frac{m_{\tilde{t}_1}^2-m_{\tilde{t}_2}^2}{4}+m_t^2+O(M_Z^2)\right)
\end{equation}
$\theta_{\tilde{t}}$ is the mixing angle of the stops. The
$\tilde{t}_2$ coupling is obtained through $\tilde{t}_1
\leftrightarrow \tilde{t}_2$. The non mixing term $m_t^2$ {\em
adds} up with the top contribution, whereas the mixing term
interferes {\em destructively} with the top. For large mixing with
large enough gap between the two stops masses  this means that a
reduction in $gg \to h$ occurs but accompanied with a more modest
increase in the $h \to \gamma \gamma$ due to the $W$ loop. Of
course the $Br(h \to \gamma \gamma)$ can be much more efficiently
increased if a drop in $h \to b \bar b$ occurs as within the
BMSSM.
Therefore we see that by letting light stops jump into the game
and keeping a ratio in the $\gamma\gamma$ channel higher than the
standard model, the correlations between the different channels
will change.

\begin{figure}[h!]
\begin{center}
\includegraphics[scale=0.3,trim=0 0 0 0,clip=true]{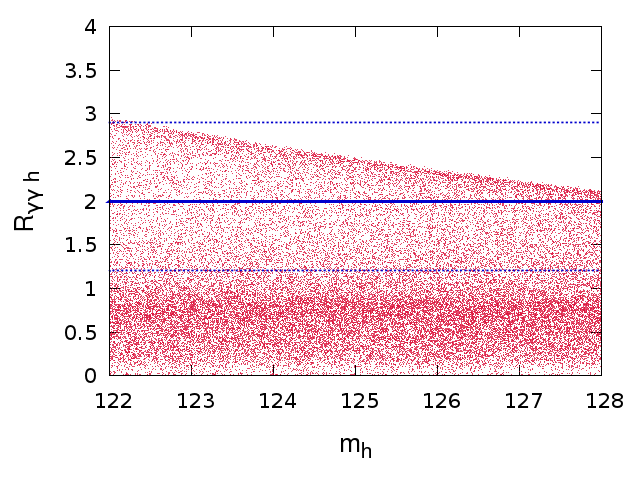}
\end{center}
\caption{\label{fig:h_Rgam2} {\em Allowed region in the plane
$m_h,\Rgam$. The blue line represents the ATLAS best fit for the
signal strength, and the dotted lines are the one sigma deviations
from this value in model B with maximal mixing and  with
$m_{\tilde{t}_2}=600$ GeV.}}
\end{figure}

\begin{figure}[h!]
\begin{center}
\begin{tabular}{cc}
\includegraphics[scale=0.3,trim=0 0 0 0,clip=true]{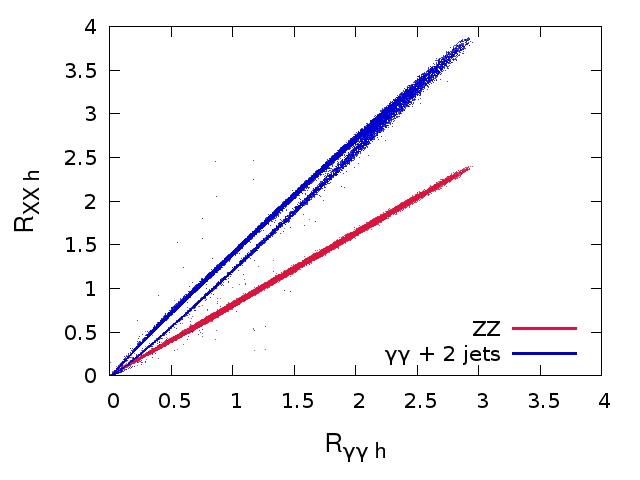}&
\includegraphics[scale=0.3,trim=0 0 0 0,clip=true]{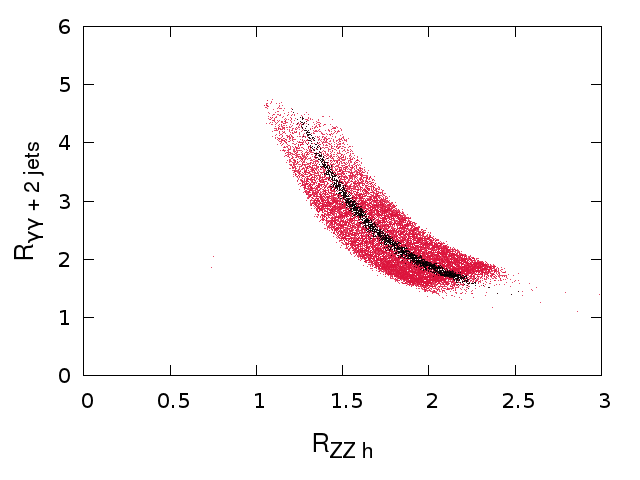}
\end{tabular}
\end{center}
\caption{\label{fig:corrstop} {\em Left panel: correlations
between $\Rgam$, $\Rz$ and $R_{\gamma\gamma+\text{2 jets}}$ for
$122< m_h < 128$ GeV in the maximal mixing scenario of model B with
$m_{\tilde{t}_2}=600$ GeV. Right panel is a subset after imposing
$\Rgamh=2.0\pm10\%$ (points in red) and $\Rgamh=2.0\pm1\%$ (points
in black) in the plane $\Rz$ and $R_{\gamma\gamma+\text{2 jets}}$
in model B scanning in the range $m_{\tilde{t}_2}\in[300,1000]$
(GeV)}}
\end{figure}

We first note, see fig.~\ref{fig:h_Rgam2}, that in the maximal
mixing case $\sin^2(2\theta_{\tilde{t}})=1$ and with
$m_{\tilde{t}_2}=600$ GeV, $\Rgam$ is reduced somehow compared to
the non mixing case of model A, however one still obtains
enhancements of a factor $2$ (and more) compared to the SM.
However, now the $\gamma\gamma+\text{2 jets}$ can be much higher
than the $\gamma\gamma$ channel, whereas previously we had
$R_{\gamma\gamma+\text{2 jets}}=1.5$ for $\Rgam=2$, now for the
same value of $\Rgam$ $R_{\gamma\gamma+\text{2 jets}}=2.5$, see
fig.~\ref{fig:corrstop}. Moreover the weight between
$R_{\gamma\gamma+\text{2 jets}}$ and $R_{ZZ}$ has been inverted,
we now have $R_{\gamma\gamma+\text{2 jets}} > R_{ZZ}$.  Scanning
over $m_{\tilde{t}_2}$ from 300 GeV to 1 TeV will open up more
possibilities for the correlations between these channels. The
results of this scan are shown in the right panel of
fig.~\ref{fig:corrstop}. For example imposing that
$\Rgam=2.0\pm10\%$ one can obtain $R_{\gamma\gamma\text{ + 2
jets}}=3.8$ together with $\Rz=1.3$. We can therefore recover
values that correspond to the best fits for these observables
obtained by the two collaborations. We stress again that this is
illustrative and shows how much flexibility in the model can be
introduced. While in the case of no-mixing in the stop sector all
channels seemed to have nearly trivial correlations, raising the
mixing will in most cases raise the $\gamma\gamma\text{ + 2 jets}$
channel compared to the $\gamma\gamma$ channel, and also decrease
the $ZZ\rightarrow 4l$ channel with respect to the $\gamma\gamma$
one.

\subsection{Prospects for other signals}
Although an unambiguous signal refuting the SM would be, in the
case where the signal at $m_h=125$ GeV were confirmed, a precise
determination of the signal strength above (or below) the SM
expectation, such a precision may require some time. At the same
time as the luminosity increases other channels and signatures may
become sensitive in corroborating the signals with $m_h \sim 125$
GeV. These channels could either be other channels where  the same
Higgs with mass 125 GeV takes part or channels affecting  the
other Higgses of the model. In the first case, the other allowed
decay modes are $\oo{\tau}\tau$ and $\oo{b}b$ final state, however
if the trend towards an increase in the $2\gamma$, $ZZ$ and
$2\gamma + 2 jets$ is reinforced requiring a reduced $hb\bar b$
(and consequently $h\oo{\tau}\tau$) in the BMSSM, the $\oo{\tau}\tau$
and $\oo{b}b$ whose current sensitivity in the SM is quite low will
require substantial increase in the luminosity.

To pursue this investigation about the prospects of signals in
other channels, we keep for the sake of illustration those models
compatible with
\begin{equation}
 1.2<\Rgam<2.9\qquad\&\qquad 0.5<\Rz<2.4,
 \label{eq:signal}
\end{equation}
which is the one sigma band obtained by the ATLAS collaboration
and show the different $R_{XX}^{{\rm exclusion}}$. Again,
eq.~\ref{eq:signal} is an arbitrary choice, taken for the sake of
concreteness. One should keep in mind that as more data is
collected, this requirement will
become stronger or perhaps even totally irrelevant.\\

\subsubsection{Non-mixing scenario}
\begin{figure}[h!]
\begin{center}
\includegraphics[scale=0.3,trim=0 0 0 0,clip=true]{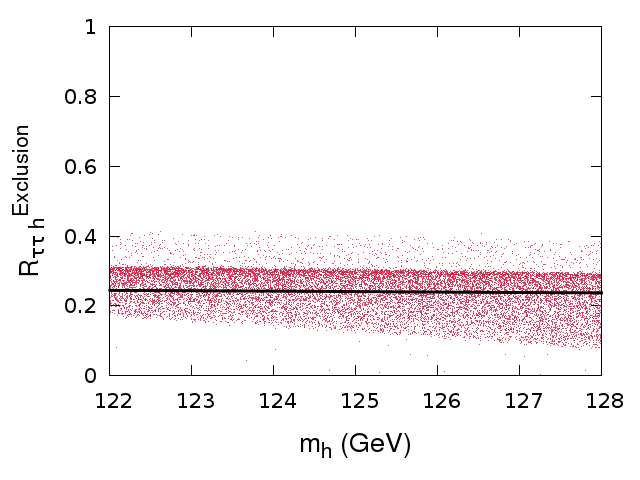}
\end{center}
\caption{\label{fig:htau} {\em Discovery perspective for the channel $h\rightarrow\oo{\tau}\tau$. The line is  black corresponds to the SM.}}
\end{figure}

We start with the $\oo{\tau}\tau$ signal of $h$. We see in fig. \ref{fig:htau} that $R_{h\tau
\tau}^{{\rm exclusion}}$ is always below $0.4$, with a
concentration below $0.3$ that corresponds also to the expectation from a SM Higgs, therefore a luminosity in excess of $30
{\rm fb}^{-1}$ is needed in the most favourable cases. Most cases
will require much more luminosity, up to 500 fb$^{-1}$ in the worst case. Incidentally we note that this channel, despite the reduced
$h \oo{\tau}\tau$ coupling, can be above that of the SM, which shows that a reduced $h\oo{\tau}\tau$ does not mean a large drop in the $\oo{\tau}\tau$
branching ratio, moreover the production cross section can be larger than in the SM.\\

\begin{figure}[h!]
\begin{center}
\begin{tabular}{cc}
\includegraphics[scale=0.3,trim=0 0 0 0,clip=true]{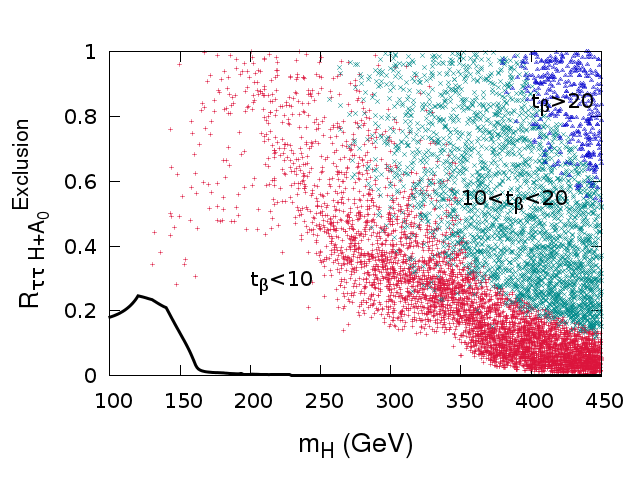}&
\includegraphics[scale=0.3,trim=0 0 0 0,clip=true]{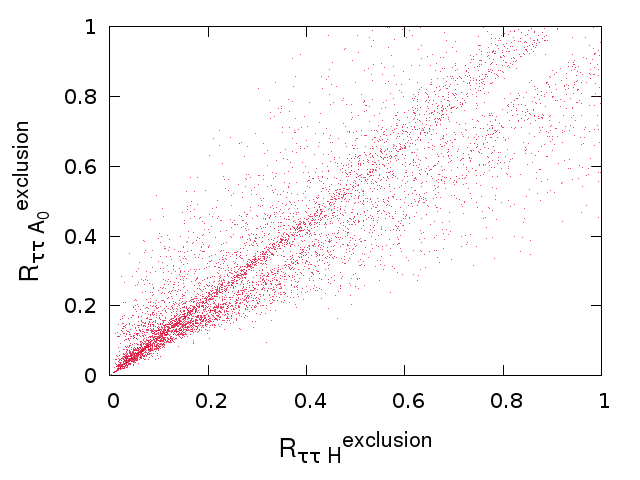}
\end{tabular}
\end{center}
\caption{\label{fig:other2} {\em Discovery perspective in the
$\oo{\tau}\tau$ channel through the heavier Higgses: in the case
where $A^0$ and $H$ are degenerate within 10 GeV (left pannel) and
when  $A^0$ and $H$ are not degenerate (right panel). In the
latter the correlations between the two signal is shown. In the
panel on the left, the SM case is shown in black. The different
shades for the BMSSM correspond (from left to right) to cases with
$\tb<10$(red) , $10<\tb < 20$ (green) and $\tb>20$(blue).  }}
\end{figure}
Would the other Higgses be more sensitive? The answer  can be
drawn from fig.~\ref{fig:other2}. Some scenarios can be probed
with little increase in the present luminosity. Generically, high
$\tb$ ($\tb>20$) will be probed within the next 30 fb$^{-1}$,
while low $\tb$ ($\tb<10$) could be quite hopeless if the heavier
Higgses are heavier than 400 GeV. We find that
$R^\text{exclusion}>0.9$ are reached in cases where $A^0$ and $H$
are close enough in mass to be degenerate ($|m_{A^0}-m_H|<10 $
GeV), yielding thus a single signal.   $R^\text{exclusion}>0.9$ is
reached also when the degeneracy is lifted, in which case one
expects both signals to be revealed with roughly the same
luminosity, see the correlation in  fig.~\ref{fig:other2}. Models
with $m_H \sim \ma <250$ GeV (degenerate case) show a ratio
$\RtauE>0.4$, which means that the region where the decoupling is
not complete between light and heavy Higgses could be probed with
about 30 fb$^{-1}$ . In the non-degenerate case, there is of
course a loss of a factor two, but there is still a lower limit to
the exclusion ratio in this mass range. But in many models,
$R^\text{exclusion} <0.4$ $A^0$ and $H$ will go undetected even
with a luminosity in excess of $30 {\rm fb}^{-1}$. This discussion
shows that studying the $\tau$ channel in Higgs physics is
crucial. Not only it can deliver new signals but can give
important information on the parameters of the model.
\\

\begin{figure}[h!]
\begin{center}
\begin{tabular}{cc}
\includegraphics[scale=0.3,trim=0 0 0 0,clip=true]{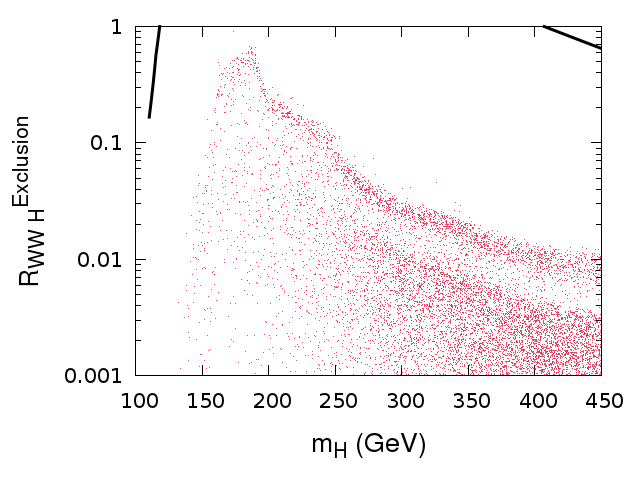}&
\includegraphics[scale=0.3,trim=0 0 0 0,clip=true]{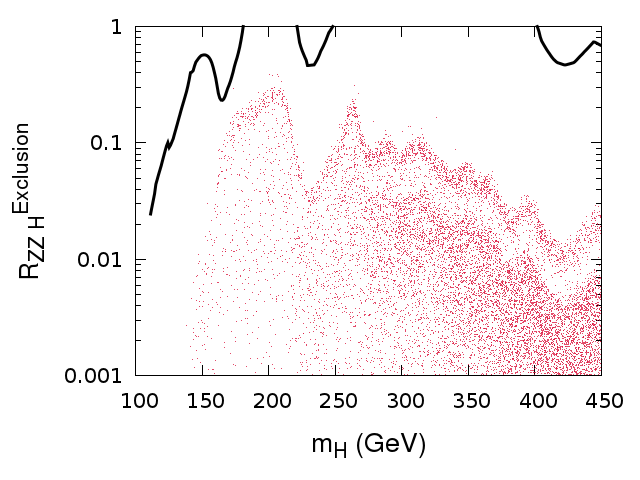}
\end{tabular}
\end{center}
\caption{\label{fig:other1} {\em Discovery perspective for other
signals : $H\rightarrow ZZ$ on the left and $H\rightarrow WW$ on
the right. No mixing or light stops are assumed. The curve in black  is the SM Higgs hypothesis. For $WW$, the curve is out of the bounding box, this confirms that for Higgs masses above the $WW$ threshold  this channel is very constraining.  }}
\end{figure}
Other channels offer little prospects, apart if $M_H \sim 180$ GeV
where the search sensitivity in the clean $WW$ and somehow also
the $ZZ$ channel is high, despite the fact that the $HWW$ is quite
small, see fig.~\ref{fig:other1}.

\subsubsection{Maximal mixing}
\begin{figure}[h!]
\begin{center}
\includegraphics[scale=0.3,trim=0 0 0 0,clip=true]{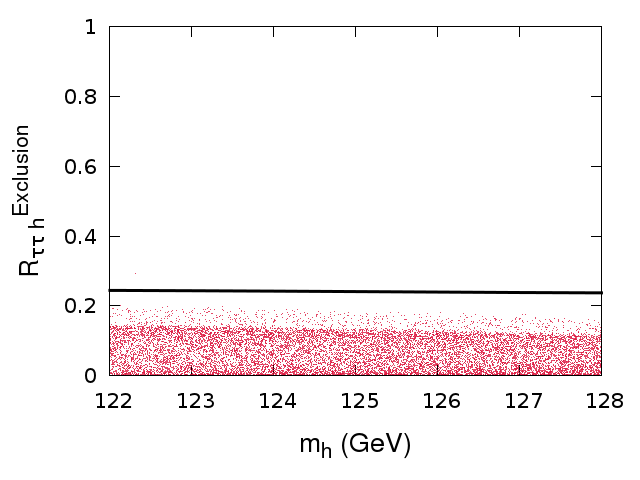}
\end{center}
\caption{\label{fig:htau2} {\em Discovery perspective for the
channel $h\rightarrow\oo{\tau}\tau$ in the maximal mixing scenario. The
line in black  represents the SM.}}
\end{figure}
In the maximal mixing case, with $m_{\tilde{t}_2}=600$ GeV, there
are few differences. The drop in $gg \to h$ is the reason behind the drop in sensitivity. Subsequently
the $\oo{\tau}\tau$ channel of $h$ will be even less sensitive, as can
be seen in fig.~\ref{fig:htau2}. $R_{h\oo{\tau}\tau}^{{\rm
exclusion}}$ is now below 0.2.
\\

As concerns the heavier Higgses, the changes are marginal compared
to the no-mixing case. The best prospects are in the $\tau$
channels and in the $WW$ channel if $m_H \sim 180$ GeV. The
corresponding figures are similar to those shown for the no mixing
case and we therefore do not display them here.

%\newpage
%%%%%%%%%%%%%%%%%%%%%%%

\section{$H$ as a signal in the $122-128$ GeV window}
\label{case_hh} As fig.~\ref{fig:pre_signal} makes clear, the
BMSSM is compatible with a scenario where it is the heavier of the
two CP even Higgses, $H$, which is in the range $122-128$ GeV and may thus be
responsible for a signal,  while the lightest Higgs $h$ has so far gone
undetected. Such possibility, even though restrained, has also
been evoked in the case of the MSSM\cite{Heinemeyer:2011aa}. We
review such a possibility in the case of the BMSSM both in a
scenario with no stop mixing and a scenario with large stop mixing
and light stops.

\subsection{Model A: No light stops, no mixing in the stop sector}
\label{case_hhmodela}
\begin{figure}[h!]
\begin{center}
\begin{tabular}{cc}
\includegraphics[scale=0.3,trim=0 0 0 0,clip=true]{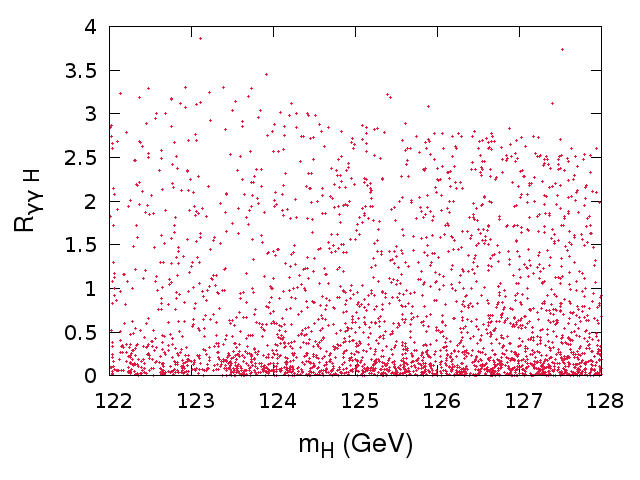}&
\includegraphics[scale=0.3,trim=0 0 0 0,clip=true]{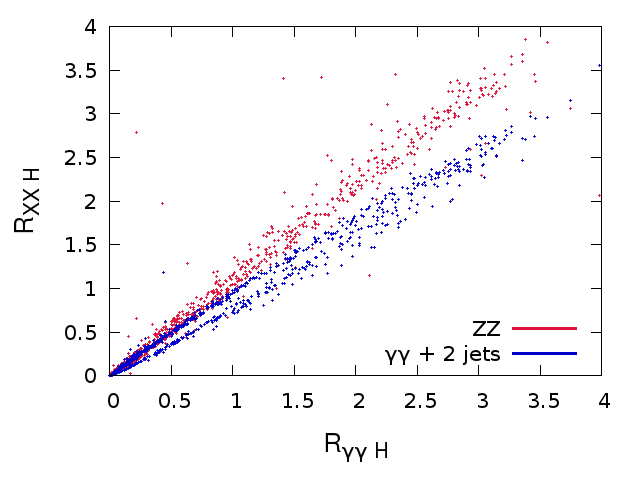}
\end{tabular}
\end{center}
\caption{\label{fig:H_Rgambis} {\em We show here the allowed
region in the plane $m_H,\Rgam$ (left panel) and the associated
correlations between $\Rgam$, $\Rz$ and $R_{\gamma\gamma+\text{2
jets}}$ for $122< m_H < 128$ GeV (right panel) in the scenario
with no-mixing .}}
\end{figure}
The statement we have just made can be made more quantitative.
Solutions with $122<m_H<128$ GeV correspond to a situation where
all three Higgses are light in the sense of being all three below
the $WW$ threshold, $m_h<120$ GeV $\ma < 160$ GeV.  We find that some features, for the signal observables,
are to a large extent similar to what we have found in the case of
$h$. In a way the $h$ and $H$ have swapped their role as to which
is the SM-like, SM-like as concerns the $VVH/h$ strength. Indeed,
this is illustrated in fig.~\ref{fig:H_Rgambis}. $\Rgam$ can still
reach values as large as 3.5, there are  correlations between
$\Rgam$, $\Rz$ and $R_{\gamma\gamma+\text{2 jets}}$ with  $\Rz >
R_{\gamma\gamma+\text{2 jets}}$ in most cases, but not all as was
the case for $122<m_h<128$ GeV. In this case, there is some spread in the
correlations between  $\Rz$ and $R_{\gamma\gamma+\text{2
jets}}$, see fig.~\ref{fig:H_Rgambis}.
\\

Let us turn to other characteristics of these BMSSM scenarios and
how they could show up in other observables.
\begin{figure}[h!]
\begin{center}
\includegraphics[scale=0.3,trim=0 0 0 0,clip=true]{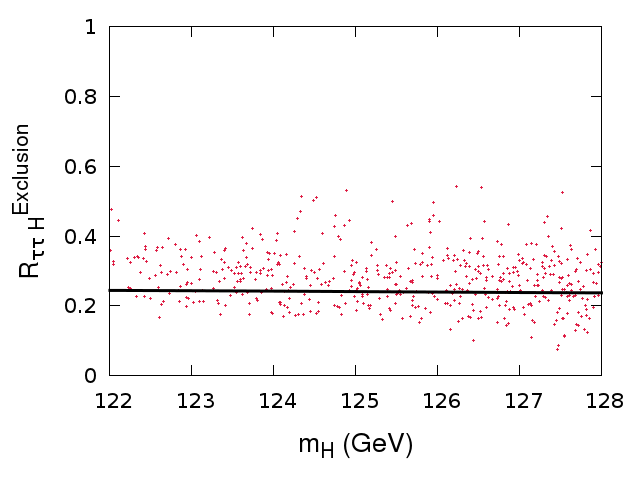}
\end{center}
\caption{\label{fig:mass_H} {\em $R_{\oo{\tau}\tau H}^{Exclusion}$ for
$122<m_H<128$ GeV. No mixing is the stop sector is assumed. The
line in black is the SM.}}
\end{figure}
Another mode where $H$ could be observed is the $H\to \oo{\tau}\tau$
channel. Fig.~\ref{fig:mass_H} suggests that prospects here might
be better than for $h$ giving a signal in the range $122<m_h<128$
GeV. Indeed, there are solutions with $R_{\oo{\tau}\tau
H}^{Exclusion}=0.5$
that would need about $20 {\rm fb}^{-1}$ to be uncovered.\\
\begin{figure}[h!]
\begin{center}
\begin{tabular}{cc}
\includegraphics[scale=0.3,trim=0 0 0 0,clip=true]{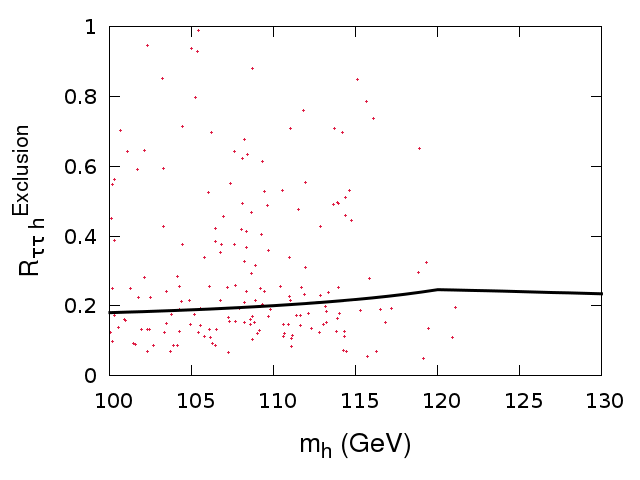}&
\includegraphics[scale=0.3,trim=0 0 0 0,clip=true]{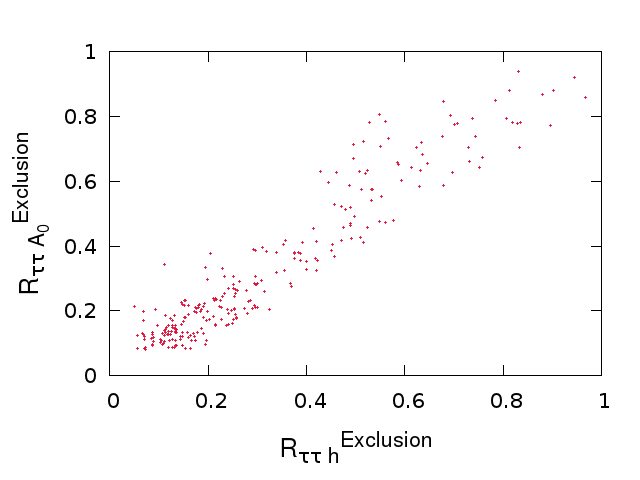}
\end{tabular}
\end{center}
\caption{\label{fig:H_other} {\em $R^\text{exclusion}_{\oo{\tau}\tau
h}$ as a function of $m_h$ for $122< m_H < 128$ GeV in the case
where $A^0$ and $h$are not degenerate within $10$ GeV (left panel).
The right panel shows the correlations between the $h$ and $A^0$
in the $\tau$ channels. No mixing in the stop sector is applied.
The line in black in the left panel represents the SM.}}
\end{figure}
Observability of the other Higgses shows, in many cases, very good
prospects, gain in the $\oo{\tau}\tau$ channels, $R_{\oo{\tau}\tau
h}^{Exclusion}>0.6$ are obtained, see fig.~\ref{fig:H_other} .
Therefore it is worth pursuing searches of $h$, for $m_h<120$ GeV
in the $\oo{\tau}\tau$ channel. $A^0$ could also be uncovered with
the same luminosity, in fact fig.~\ref{fig:H_other} shows the
correlation between $h$ and $A^0$ in the $\oo{\tau}\tau$ channel.
There, of course, remains also many situations with $R_{\tau
\tau}^{Exclusion}<0.2$ that would be difficult to decipher.
%%%%%%
%%--------

%\newpage

\subsection{Model B: a light stop with large mixing in the stop sector}
\label{case_hhmodelabis} 
\begin{figure}[h!]
\begin{center}
\includegraphics[scale=0.3,trim=0 0 0 0,clip=true]{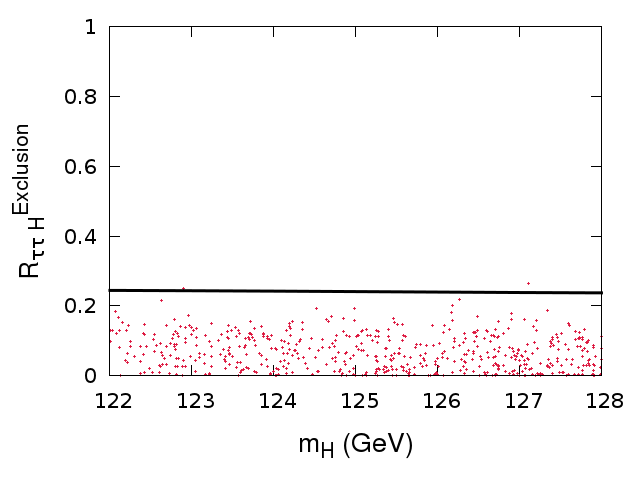}
\end{center}
\caption{\label{fig:mass_H2} {\em $R_{\oo{\tau}\tau H}$ in the same
mass $122<m_H<128$ GeV in the case of maximal stop mixing and
$m_{\tilde{t}_2}=600$ GeV. The line in black is the SM. }}
\end{figure}

We now turn to the maximal mixing case
and restrict  ourselves to $m_{\tilde{t}_2}=600$ GeV
($m_{\tilde{t}_1}=200$ GeV). Compared to the previous case without
mixing one notes that there is a reduction in $R_{\oo{\tau}\tau H}$.
This is mainly driven by  the drop in
$gg \to H$, see fig.~\ref{fig:mass_H2}, very low values are also due to quite small $h \tau\bar \tau$ couplings.
\\

\begin{figure}[h!]
\begin{center}
\begin{tabular}{cc}
\includegraphics[scale=0.3,trim=0 0 0 0,clip=true]{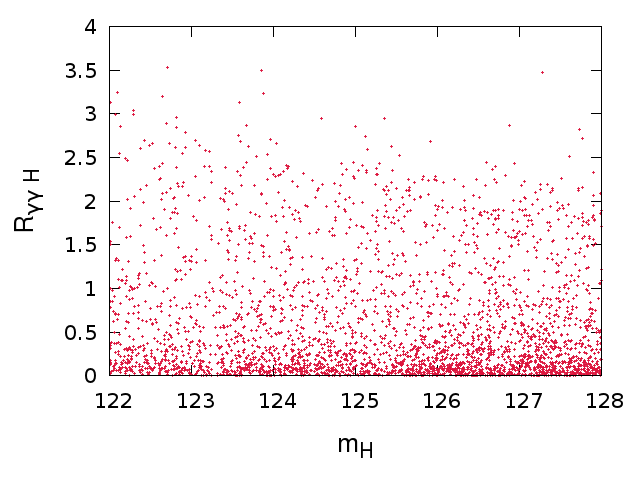}&
\includegraphics[scale=0.3,trim=0 0 0 0,clip=true]{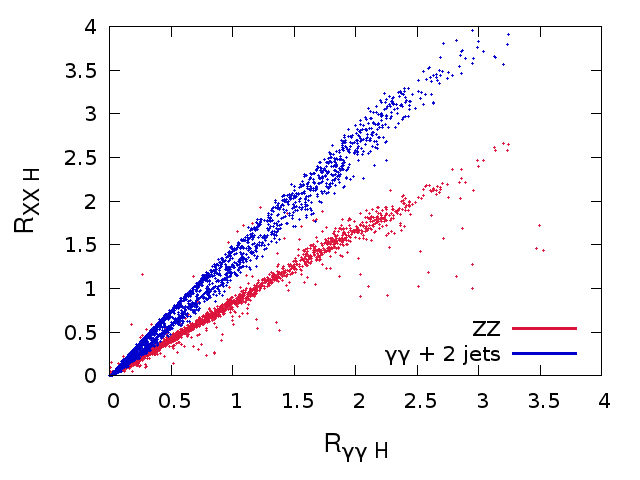}
\end{tabular}
\end{center}
\caption{\label{fig:H_Rgam2bis} {\em We show here the allowed
region in the plane $m_H,\Rgam$ (left panel) and the associated
correlations between $\Rgam$, $\Rz$ and $R_{\gamma\gamma+\text{2
jets}}$ for $122< m_H < 128$ GeV (right panel) with maximal mixing
and $m_{\tilde{t}_2}=600$ GeV.}}
\end{figure}

The most noticeable change is
the correlation between $\Rz$ and $R_{\gamma\gamma+\text{2
jets}}$, see fig.~\ref{fig:H_Rgam2bis}. We now easily find
$R_{\gamma\gamma+\text{2 jets}} > \Rz$. The spread in this
correlation has increased. One can find scenarios with $R_{ZZ}<1$
even for $\Rgam>2$. For $\Rgam \sim 2$, $R_{\gamma\gamma+\text{2
jets}}> 2$ is attained.
\\

The visibility of $A^0$ and $h$ is little affected by the stop mixing. Our conclusions are little changed. Again it is very
important to pursue the search in the $\oo{\tau}\tau$ channel.
\newpage

\section{Conclusion}
\label{sec_conclusions} Despite the fact that no sign of
supersymmetry has been found so far, the BMSSM framework is a very
efficient set up that extends the realm of the MSSM in a most
natural way as concerns the realisation of the Higgs. In the MSSM
framework there is some tension with naturalness for a Higgs mass
of $125$ GeV that requires heavy stops, in the BMSSM this is not an
issue. Although one must exercise extreme care with the so called
tantalising hints for a Higgs signal around this mass, 125GeV, it
is extremely important to scrutinise the properties of the Higgs
with such a mass in many models, in particular the BMSSM which
represents an effective implementation of a variety of
supersymmetric models having the same field content as the MSSM.
The tantalising hints have come also with the temptation, even if
premature,  of reading from  the results of ATLAS and CMS, despite
the uncertainty of the measurements, that the signals in the
inclusive $2\gamma$ channel, the $2\gamma + \; {\rm jets}$ and
perhaps in the $ZZ \to 4l$ to be higher than what is expected from
the SM. Such scenarios are practically  impossible to attain in
the MSSM, a possibility that has been entertained would make the
naturalness argument even more excruciating. It is therefore very
important to find out whether some configurations, especially
those leading to enhancements in these most important channels can
be realised in the BMSSM. As important is to find out how these
enhancements or signals are correlated and how different kinds of
correlations can be realised. We have shown that a vanilla BMSSM
where stops are at very moderate masses and with little mixing
easily allows enhancements in all these channels for $m_h \sim
125$ GeV with the constraint that the rate $ZZ \to 4l$  would
generally be higher than the rate $\gamma \gamma \;+ 2 {\rm
jets}$. A light stop with large mixings in the stop sector offers
more possibilities especially as concerns correlations between
these three important channels. Our study also reveals that
although it is easier to have such realisations work for the
lightest Higgs of the BMSSM, solutions where it is the heaviest
Higgs that has a mass around 125 GeV also exist. Once a signal at
$125$ GeV has been confirmed a better measurement of the rates, in
particular the $2\gamma$ (inclusive and exclusive) as well as the
$4l$ would narrow considerably the parameter space of the BMSSM.
At the same time as more precision is  achieved and more
luminosity is gathered one can constrain the models through the
other Higgses (those outside the 125 GeV window) but also through
other channels of the Higgs at 125 GeV. Our study reveals that in
both cases it is crucially important and telling to investigate
the $\oo{\tau}\tau$ channel. We have not folded in the possible
constraints from flavour physics and dark matter as we have argued
that this introduces some model dependencies (including those from
cosmology) but it is clear now that we have entered a fascinating
era. The study we have conducted is an example which shows that
even before any new direct signal of New Physics is discovered,
the study of the Higgs, once confirmed, will give important clues
on the New Physics. We eagerly await more data and analyses from
the experiments and we urge, once more, our colleagues to provide
as much information as possible on the data so that one can gain
access to the different individual subchannels that make up an
inclusive channel.

\renewcommand{\thesection}{\Alph{section}}
\setcounter{section}{0}
\newpage

\section{Addendum: Tevatron and the $\bar{b} b$ channel}
\begin{figure}[h!]
\begin{center}
\includegraphics[scale=0.3,trim=0 0 0 0,clip=true]{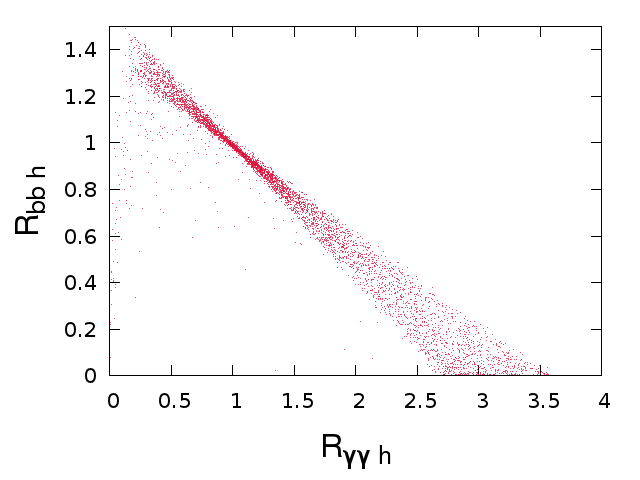}
\end{center}
\caption{\label{fig:tev_bb} {\em Correlation between diphoton
channel ($R_{\gamma\gamma}$) and the $VH\to V\oo{b}b$
($R_{\bar{b}b}$) in the non-mixing scenario.}}
\end{figure}

While this work was being finalised, the Tevatron Collaborations
released new analyses\cite{tevatron_10fb} pointing out to a
possible  signal in $VH\to V \bar b b$ channel with a rate that
could be  compatible with the Standard Model expectation and with
a mass that could correspond to where the excesses are seen at the
LHC. This would seem at first sight to disfavour a scenario where
$g_{h\bar{b}b}$ is very much reduced. However,  one must keep in
mind that since the decay $H \to \bar{b}b$  dominates for
$m_h=125$ GeV, a suppression of the coupling by a factor two does
not imply a suppression of the branching ratio by a factor two.
The suppression in much more modest and there can still be a
significant enhancement of the diphoton channel without
suppressing too much the $VH\to V\bar{b}b$ channel. It must be
stressed that a more precise  measurement of the latter process
would really be helpful. Indeed, there exists also a correlation
between the diphoton (inclusive) channel and this channel, as
shown in fig \ref{fig:tev_bb}.

% \bibliography{biblio}

\end{document}